\documentclass[epjCONF,columns]{svjour}
\usepackage{graphics}
\usepackage[varg]{txfonts} 
\usepackage[latin1]{inputenc}
\session-title{Conference Title, to be filled}
\begin{document}
\title{B-tagging at CMS}
\author{Cristina Ferro\inst{1}\fnmsep\thanks{\email{Cristina.Ferro@cern.ch}}}
\institute{IPHC Strasbourg.}
\abstract{
The identification of $b$ jets is a crucial issue to study and characterize various channels like top quark events and many new physics scenarios. Different b-tagging techniques are defined in CMS which benefit from the long life time, high mass and large momentum fraction of the b-hadron produced in b-quark jet. Efficient algorithms have been developed based on the measure of b-hadron secondary vertex or on tracks with a large impact parameter. Data collected in $pp$ collisions at $\sqrt{s}$=7TeV in 2011 are used to estimate both the b-tagging efficiency and the mistag rate from light flavor jets. 
} 
\maketitle
\section{Introduction}
\label{intro}

The b-tagging algorithms in CMS mainly rely on the long life time, high mass and large momentum fraction of b hadrons produced in b-quark jets, as well as on the presence of soft leptons from semi-leptonic b decays\cite{ref:BTV_11_001}.
Due to the high instantaneous luminosity during the 2011 data taking, the number of collision taking
place in the same bunch crossing (pileup events) is of the order of 5 to 11 on average.
\begin{figure}[!ht]
\centering 
\resizebox{0.9\columnwidth}{!}{
\includegraphics{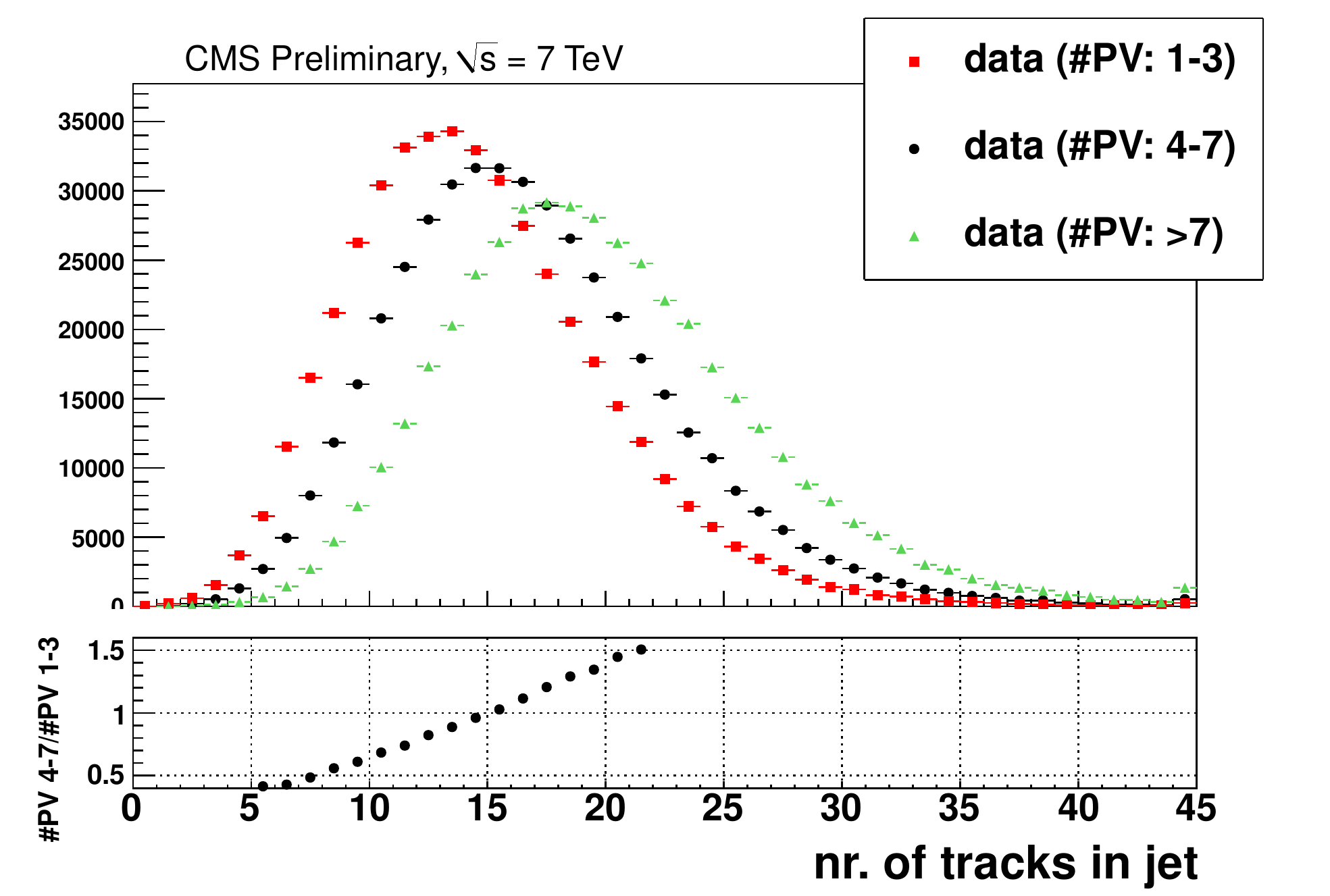} }
\caption{Number of tracks associated to a jet without any selection cut.}
\label{TrackMulty}
\end{figure}
\begin{figure}[!ht]
\centering 
\resizebox{0.9\columnwidth}{!}{
\includegraphics{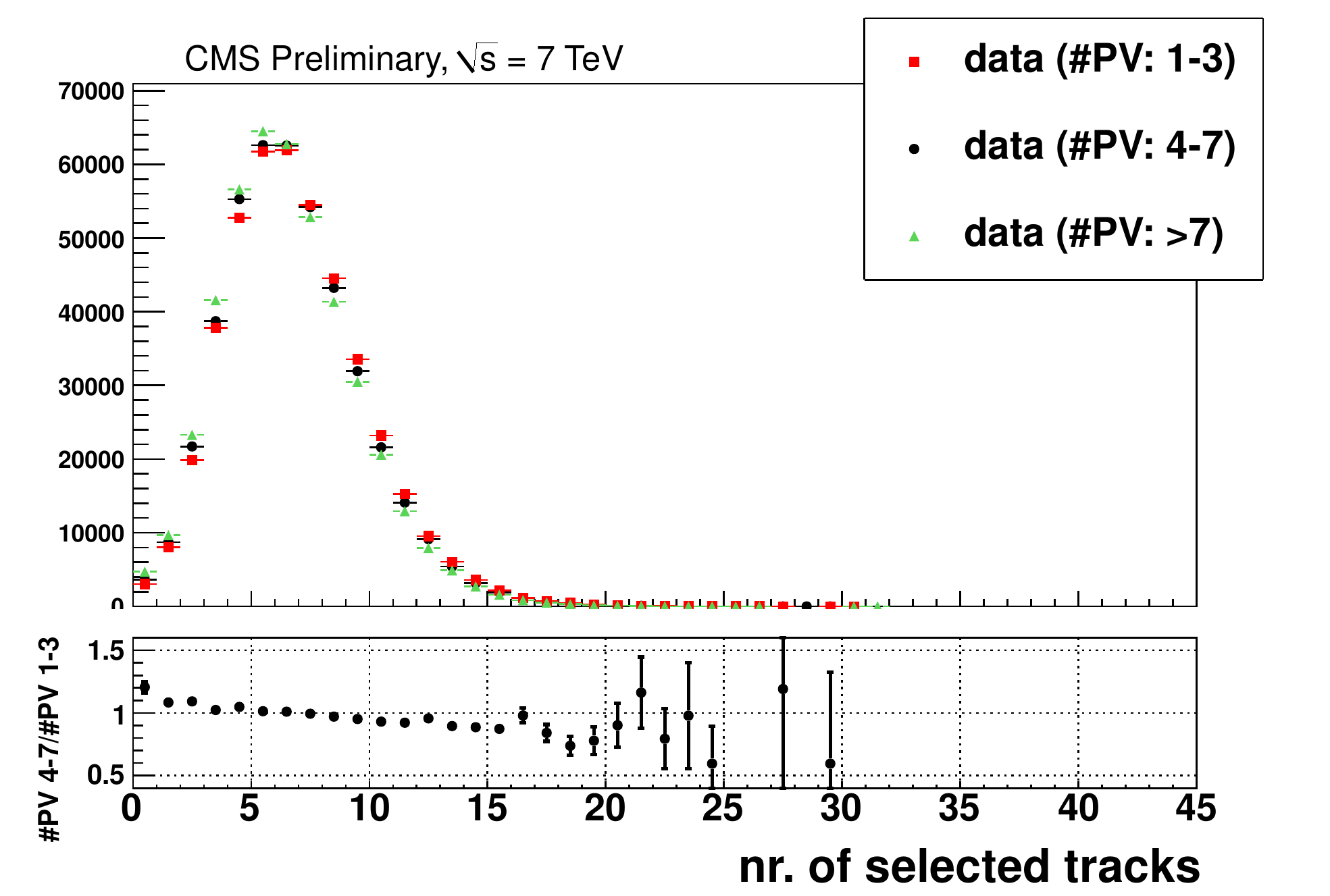}}
\caption{Number of tracks associated to a jet after selection cuts.}
\label{Track_Selected}
\end{figure}

The presence of pileup increases the track multiplicity in the events, as we can see in Fig.(\ref{TrackMulty}). This is why a special selection of the tracks was applied in order to remove the tracks originating from pileup\cite{ref:BTV_11_002}. 
In Fig.(\ref{Track_Selected}), the number of tracks passing the selection cuts shows a smaller pileup dependence.

\section{The b-tagging observables}
\label{sec2}
The b-tagging algorithms and their study are based on the measure of three main variables: the impact parameter significance of the tracks, the position of the secondary vertex, and the transverse momentum of the muon relative to the jet direction. In the following a brief description of these variable is presented.\\
\subsection{The impact parameter significance}
\label{sec2.1} 
\begin{figure}[!ht]
\centering
\resizebox{0.9\columnwidth}{!}{
\includegraphics{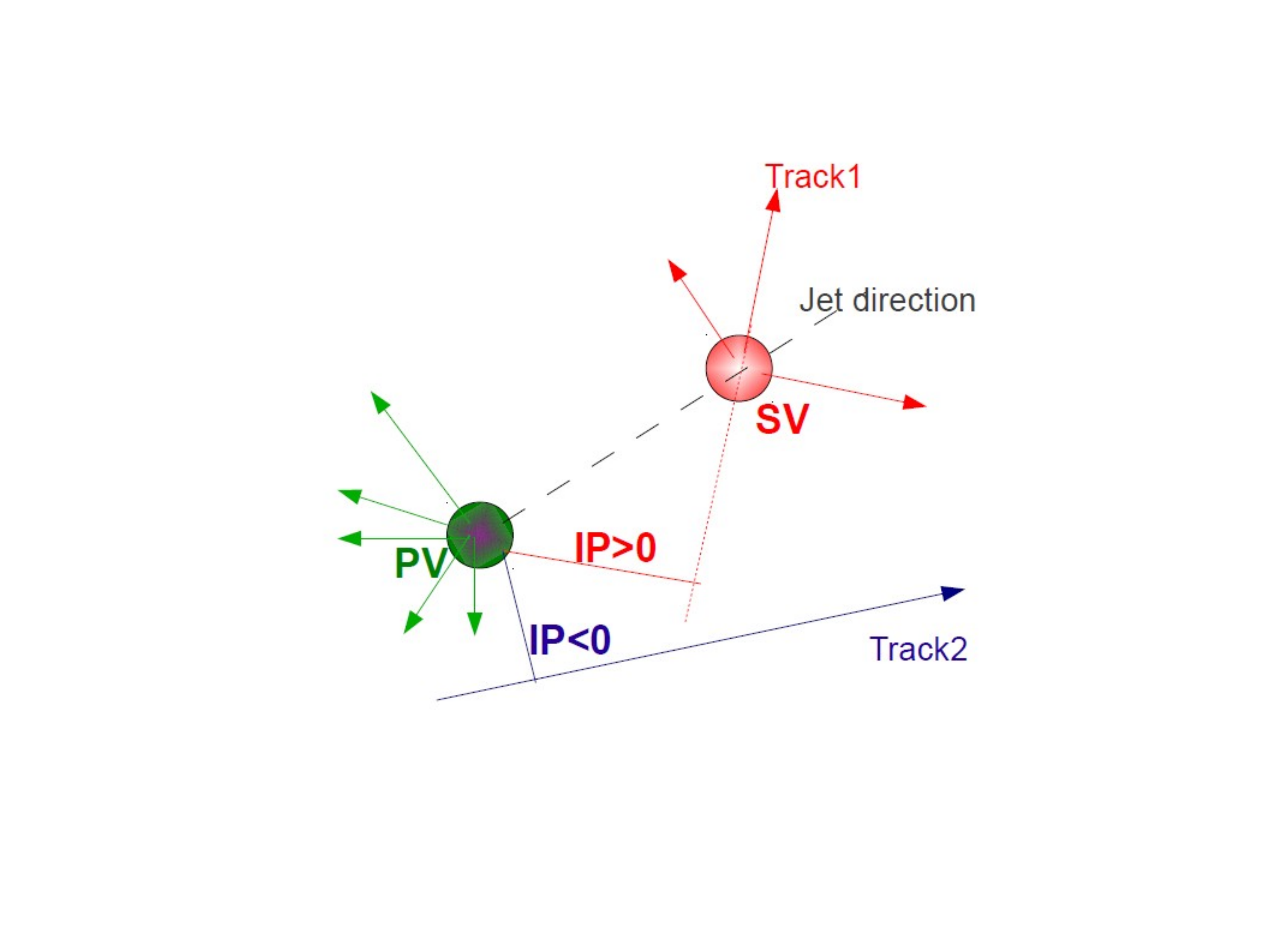}}
\caption{Geometric meaning of the impact parameter significance.}
\label{IP_def}
\end{figure}

The impact parameter (IP) is defined as the distance between the track and the primary interaction vertex (PV)
at the point of closest approach. The IP is positive (negative) if the track is produced downstream (upstream) with respect to the PV along the jet direction (Fig.(\ref{IP_def})).
The IP is calculated in 3 dimensions thanks to the good x-y-z resolution provided by the pixel
detector.
An important features of the IP is that it is Lorentz invariant and due to the b-hadron lifetime the typical IP scale is set by c$\tau \sim$480 $\mu$m. In practice, the impact parameter significance IP/$\sigma$(IP) is used in order to take into account resolution effects.Thanks to the long lifetime of the b-hadrons the IP from b-jets is expected to be mainly positive, while for the light jets it is almost symmetric with respect to zero (Fig.\ref{algo}).
\subsection{The secondary vertex}
\label{sec2.2} 
Thanks to the high resolution of the CMS traking system, it is possible to directly reconstruct the secondary vertex, the point where the b hadron decays (Fig.(\ref{IP_def})). The vertex reconstruction is performed using the adaptive vertex fitter. The resulting list of vertices is then subject to a cleaning procedure, rejecting SV candidates that share 65$\%$ or more of their tracks with the PV. 
\subsection{The transverse momentum of the muon}
\label{sec2.3} 
Semileptonic decays of b hadrons give rise to b jets that contain a muon with a branching ratio of about 11$\%$, or 20$\%$ when b$\rightarrow$c$\rightarrow$l cascade decays are included.
This is why the reconstructed muons inside a jet are used to study the performance of the lifetime-based tagging algorithms. The muons are
seeded from the CMS muon chambers, and are then linked to tracks found in the tracking
system to form global muons. The CMS muon system is able to measure muons with high acceptance resolution and efficiency.
\section{B-tagging algorithms}
\label{sec3} 
\begin{figure}[!ht]
\centering 
\resizebox{1.1\columnwidth}{!}{
\includegraphics{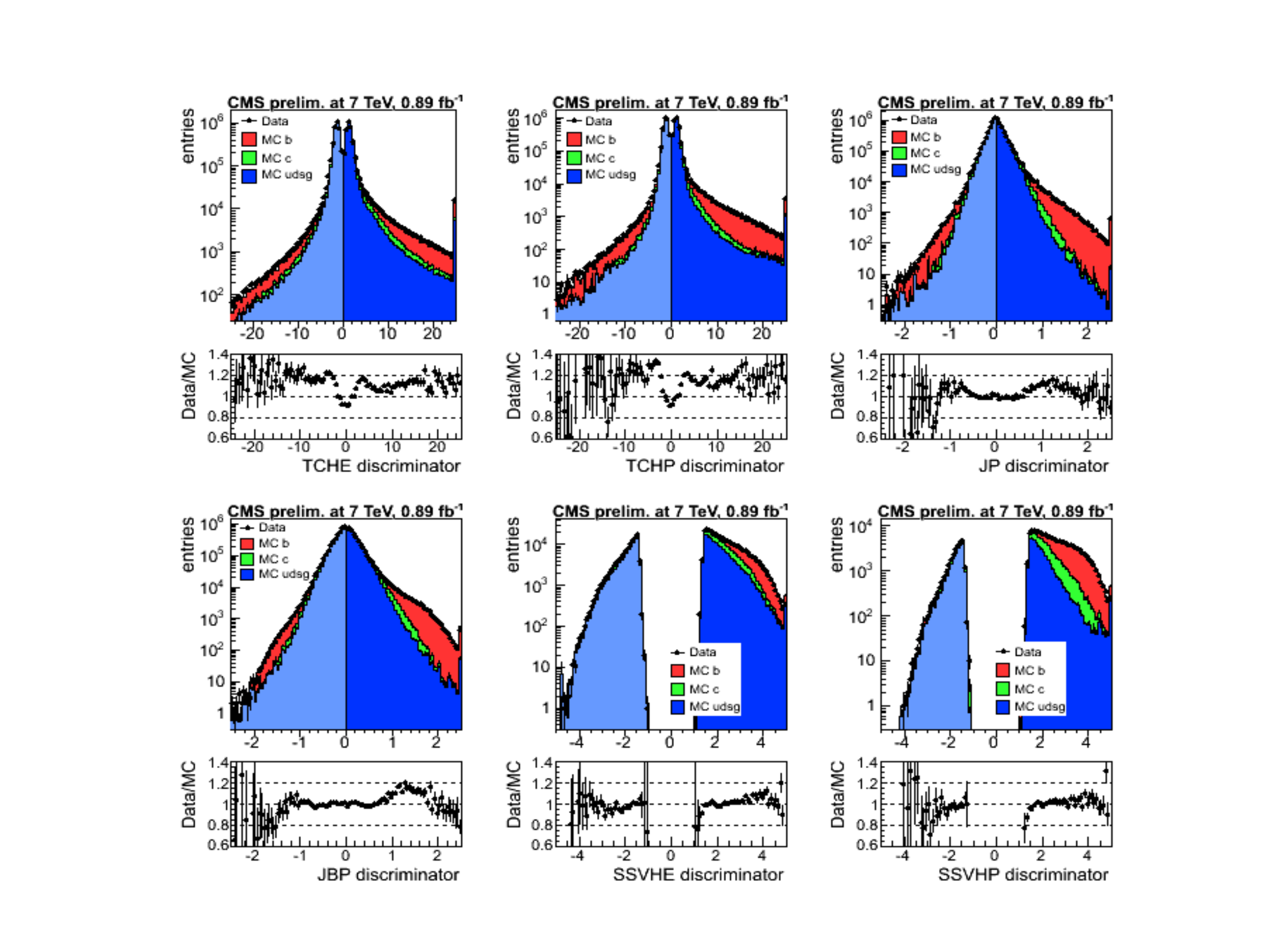}}
\caption{Discriminators for: $Top$ $left$ Track Counting High Efficiency (IP/$\sigma$(IP)), $center$ Track Counting High purity (IP/$\sigma$(IP)), $right$ JetProbability, $Bottom$ $left$ JetBProbability, $center$ Simple Secondary Vertex High efficiency, $right$ Simple Secondary Vertex High purity.}
\label{algo}
\end{figure}

Severla b-tagging algorithms are used in CMS \cite{ref:BTV_11_001}, \cite{ref:BTV_11_002}.\\
The output of each algorithm is a $discriminator$ value on which the user can cut on to select
different regions in the efficiency versus purity phase space. 
In Fig.(\ref{algo})these discriminators are presented.

\begin{itemize} 
\item The track counting algorithm identifies a b-jet if there are at least N tracks with a significance
of the impact parameter above a given threshold. The tracks are ordered in decreasing IP/$\sigma$(IP) and the
discriminator is the impact parameter significance of the Nth track .
To get an high b-jet efficiency we can use the IP/$\sigma$(IP) of the second track (TCHE),
to select b-jets with high purity the third track is the better choice (TCHP).
\item The Jet Probability algorithm relies on the IP/$\sigma$(IP) measurement of all tracks in a jet. One can use the negative tail of the IP/$\sigma$(IP) distribution to extract the probability density function (PDF) for tracks not coming from b/c-jets. By integrating on the PDF, we can compute the $probability$ for tracks to originate from the PV. Then combining the probability of the tracks we can assign to the jet a probability to come from the PV. 
The JetBprobability is then defined in a similar way but giving more weight to the four most displaced tracks.

\item Soft-Lepton tagging algorithms rely on the properties of muons or electrons from semileptonic b-decay. Due to the large b-quark mass, the momentum of the muon transverse to the jet axis, $p_T^{rel}$ , is larger for muons from b-hadron decays than for muons in light flavor jets.
\item Secondary Vertex tagging algorithms rely on the reconstruction of at least one secondary vertex.
The significance of the 3D flight distance is used as a discriminating variable. Two variants based on
the number of tracks at SV are considered: N$\geq$2 for $high$ $efficiency$ (SSVHE), and Ntr$\geq$3 for $high$ $purity$ (SSVHP) [2].

The $combined$ $secondary$ $vertex$ algorithm includes this information and
provides discrimination even when no secondary vertices are found. The mass of reconstructed charged
particles at the secondary vertex is used to measure the b-tagged sample purity.
\end{itemize}
\section{Performance of the taggers}
\label{sec4}
\begin{figure}[!ht]
\centering 
\resizebox{0.6\columnwidth}{!}{
\includegraphics{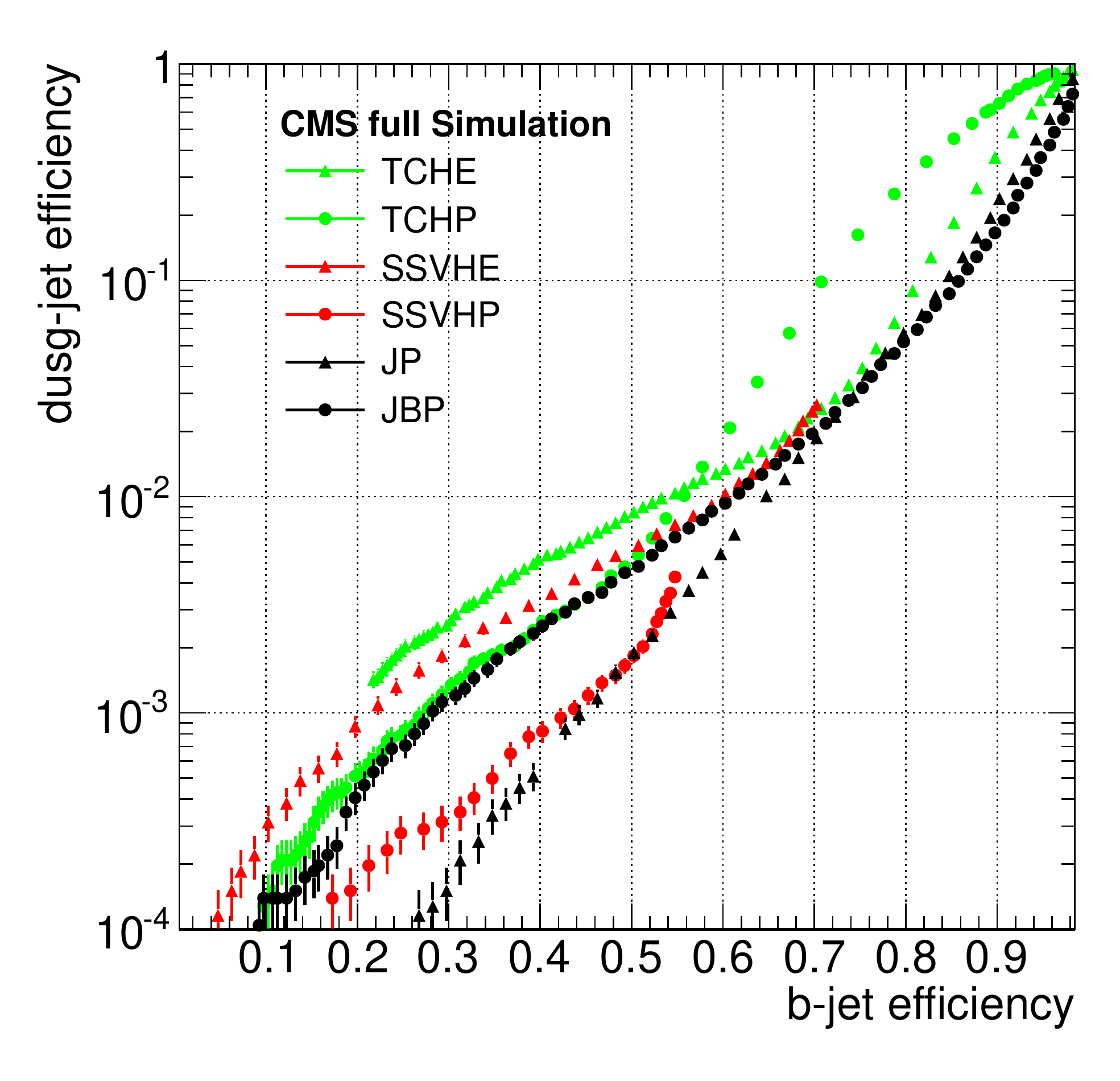}}
\caption{Performance of all b-taggers obtained on the simulated QCD events.
The performance are shown as udsg jets tagging efficiency versus b-jets tagging efficiency.}
\label{performance}
\end{figure}

\begin{figure}[!ht]
\centering 
\resizebox{0.6\columnwidth}{!}{
\includegraphics{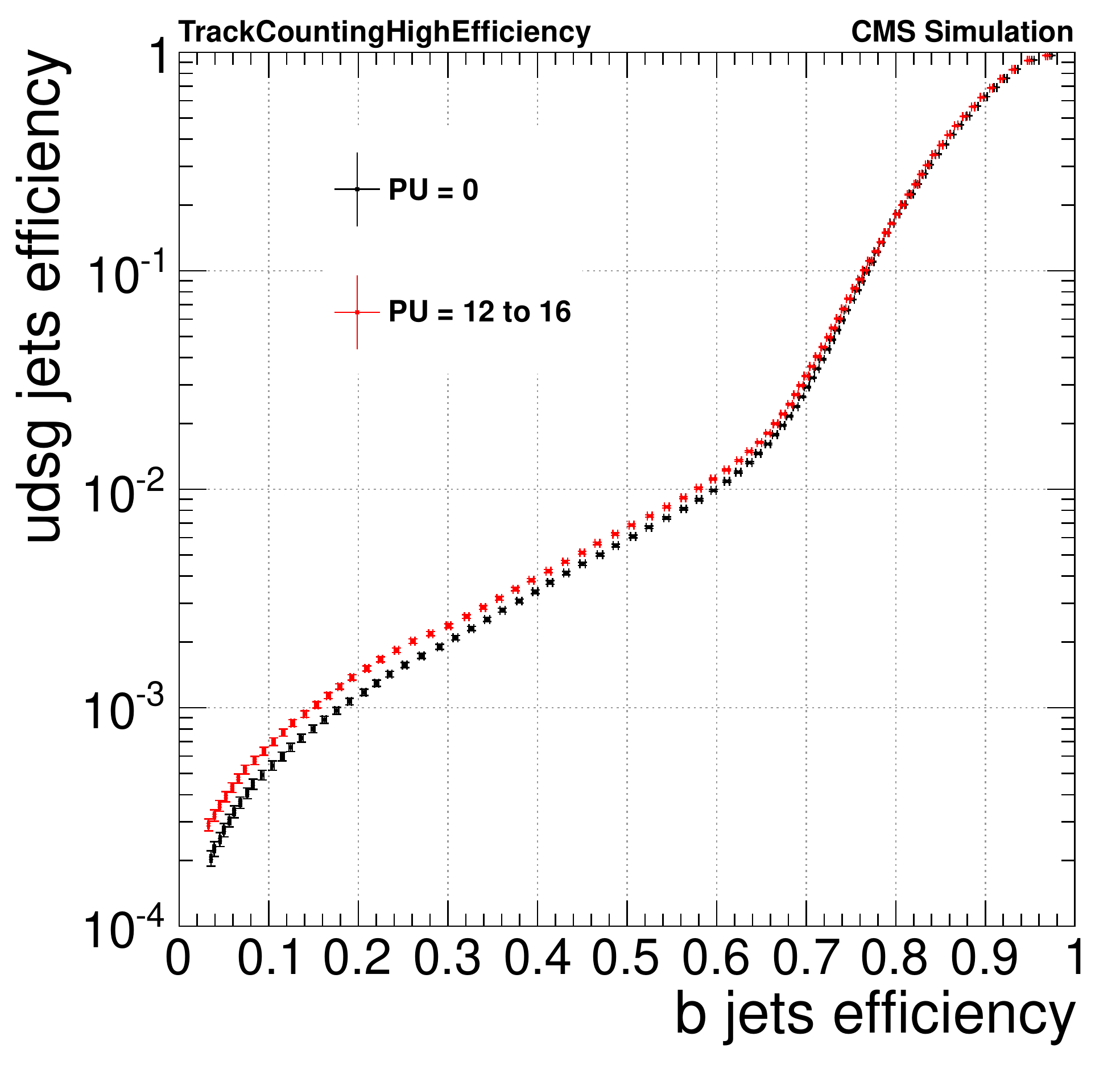}}
\caption{Light flavor mistag efficiency versus b-tagging efficiency for
different pileup scenario, for the TCHE tagger.}
\label{PU_performance}
\end{figure}

Varying the cuts on the discriminator, we
obtain different efficiency of the taggers.
We establish standard operating points
as, $loose$ (L), $medium$ (M), and
$tight$ (T), being the value at which the
tagging of udsg jets is estimated from MC
to be 10$\%$, 1$\%$, or 0.1$\%$, respectively,  for jet transverse momentum of about 80 GeV.
In Fig.(\ref{performance}) the performance for
different taggers are shown.
In Fig.(\ref{PU_performance}) the effects of the pileup
on the performance of the TCHE tagger is
presented. Thanks to the good selection on tracks the performance of the taggers are not compromised by the pileup events.
\section{Physics results}
\label{sec5}
Many measurements have been obtained using the b-tagging
algorithms at $\sqrt{s}$ = 7 TeV. Some of them used the b-tagging algorithms already at trigger level\cite{ref:TOP-11-007}. Indeed, at trigger level, the b-quark candidates can be selected if they
have at least one or two tracks with a 3D impact parameter significance above a given threshold.
The motivation for applying b-tagging in the trigger is a reduction of the trigger rates, while keeping the signal efficiency high at the same time.
The typical rate reduction is a factor of 5-10.
In the following a list of the main 2011 physics results obtained thanks to the b-tagging algorithms is presented:
\begin{itemize}
\item B-PHYSICS:
\begin{itemize}
\item Inclusive production cross section of b-jets\cite{ref:4}.
\end{itemize}
\item EW PHYSICS:
\begin{itemize}
\item Measurement of associated charm production in W final state\cite{ref:8}.
\end{itemize}
\item Top-PHYSICS:
\begin{itemize}
\item Cross-section measurement of top pair production in various final
states: dileptonic \cite{ref:9},\cite{ref:10}, \cite{ref:11}, \cite{ref:14}, lepton+jets \cite{ref:12}, all hadronic \cite{ref:TOP-11-007}.
\item Single top in t channel \cite{ref:13}.
\item Top mass measurement \cite{ref:14}.
\end{itemize}
\item New PHYSICS:
\begin{itemize}
\item Search for supersymmetry in events with b-jets and missing
transverse momentum \cite{ref:15}. 
\item Search for supersymmetry in all hadronic events \cite{ref:16}.
\item Search for an Heavy Bottom-like quark \cite{ref:17}. 
\item Search for an Heavy Top-like quark \cite{ref:18}.
\item Search for pair production of a fourth-generation t' quark in the
lepton-plus-jets channel \cite{ref:19}. 
\item Inclusive search for a fourth generation of quarks \cite{ref:20}. 
\end{itemize}
\end{itemize}

\end{document}